# World Migration Degree

## Global migration flows in directed networks

Idan Porat[1], Lucien benguigui[2]

1. The Center for Urban and Regional Studies, Technion- Israel Institute of Technology, idan.prt@gmail.com. 0542295577. 048294019
2. Solid State Institute and Physics Department, Technion - Israel Institute of Technology

**Abstract**

In this article we analyze the global flow of migrants from 206 source countries to 145 destination countries (2006-2010) and focus on the differences in the migration network pattern between destination and source counters as represented by its degree and weight distribution. Degree represents the connectivity of a country to the global migration network, and plays an important role in defining migration processes and characteristics. Global analysis of migration degree distribution offers a strong potential contribution to understanding of migration as a global phenomenon. In regard to immigration, we found that it is possible to classify destination countries into three classes: global migration hubs with high connectivity and high migration rate; local migration hubs with low connectivity and high migration rate; and local migration hubs with opposite strategy of high connectivity and low migration rate. The different migration strategies of destination countries are emerging from similar and homogenies pattern of emigration from source countries were similar network patterns were found for most of the countries. These findings, of similar behavior which creates different results is a complex phenomenon which represents the diverse nature of migration.

# 1. Introduction

Migration is an important aspect of human society. It has existed throughout history, but in modern times it takes a new form, scope and scale. Developments in transportation and, more recently, globalization have increased migration to such an extent that today it affects almost all the countries in the world (Goldin et al, 2011; Nijkamp et al, 2011; Nijkamp et al, 2012). Migration affects not only societies but also world economy, labor, mobility, environment and politics (Castles, 2010; Cohen, 1995; Koser, Laczko, 2010), and therefore, understanding its scope and dynamics is essential for understanding the world that we all live in.

Despite the intense and vast research of migration as a social and geographical phenomena and the vast literature relating to these topics, global migration network as a spatio-temporal global analysis is rather a new and limited research topic, due to shortage of reliable and comprehensive global data (Ratha and Shaw, 2007; World Bank, 2011; OECD, 2011). In recent years, data collection at a global scale of source and destination countries is evolving(Davis et al., 2013; Fagiolo & Mastrorillo, 2013; Tranos et al., 2014), and the "big picture" of world migration characteristics is beginning to clear. These studies are relating to the global patterns of migration in regional scale of and in several time periods. These studies are innovative and groundbreaking in their approach to migration; and this article contributes to this line of research whit a new analysis of bilateral migration matrix of global migration flow. This research analyze the migration pattern of 206 source and 145 destination countries and focus on the differences in the migration pattern between in and out degree.

This study is relating to the global flow of migrants from source and to destination countries between the years 2006-2010. According to World Bank data source of bilateral migration, in 2006 there were 204 million people that lived outside their home country (Ratha, et al., 2006); in 2010 this number increased to 216 million (Ratha and Shaw, 2007; updated with additional data for 71 destination countries as described in the Migration and Remittances World Bank Factbook, 2011). In the four years 2006-2010 the world experienced migration flow of 12 million people. Here we analyze this flow of migrants and focus on their source countries and destination countries migration pattern.

## 2. Migration Dynamics

Migration as a global phenomenon has increase during the last decades. During the 40 years, 1960- 2000 the number of migrants has increased from 92 million to 165 million, an 80% increase (Ozden et al 2011). In the next 10 years 2001-2010 migration has accelerated and grown to 216 million, a 30% increase (World Bank 2010). According to Ozden et al., (2011) during that period (1960-2000), migration between developed countries (north-north) and migration from developed countries to developing countries (north-south) have experienced minor changes; migration from developing countries to developed countries (south-north) have increased; and migration between developing countries (south-south) have experienced a major increase and tripled its weight.

This increase in migration also effects global migration network connectivity. According to Davis et al., (2013), at the $1960^{th}$ 50% of the world countries has more than 100 cumulative undirected stoke degree, and in the $2000^{th}$ 80% of the countries have passed this connectivity level. Therefore the level of connectivity of the global migration network is growing and more and more links between countries are formatting.

Studies of modern migration focus mainly on two approaches that are often presented together: a description of the phenomenon itself, in particular between two countries (from source to destination), and a discussion of a specific case study with its unique aspects such as differences between the characteristics of (mostly) in-migration to countries. Case studies such as the most reported migration waves between Mexico and the USA (e.g. Massey, Goldring and Durnad, 1994; Massey& Taylor, 2004), highlight three main characteristics: social capital, human capital and market consolidation - these characteristics combined with the mutual border generate the migration. Another example is Spain which has experienced a migration wave from Latin American and South American countries during the last decade (Durand and Massey, 2010). This wave has been interpreted as resulting from mutual culture, social networks, and Spanish migration policies (Sandell, 2006). Israel experienced a wave of immigration from the former Soviet Union at the beginning of the 1990s when the Soviet Union experienced political change (Alterman, 2002). It brought one million migrants (one-fifth of the country's population) in less than a decade, and then subsided.

In many cases, migration between two countries can be explained by migration system theories (Kritz et al., 1992, Castels & Miller, 2008) that analyze source and destination and their relationship, e.g. the former colonialist relationship between France and Algeria, mutual trade interests in commonwealth countries - the UK, South Africa, Australia, New Zealand - or military involvement between Korea and the USA. Middle East countries are powerful migration magnets but only from specific sources, each of them forms a close migration network separate from the global network, and with specific social characteristics (Okruhlik & Conge, 1997; Shah & Menon, 1999; Ratha & Shaw, 2007; Nasra & Shah, 1999).

According to chain migration theory, migration dynamics usually begins with a link (even a weak link) of social network between migrants and their community in the home land (De Hass, 2010a,b; Massey, 1990; MacDonald and MacDonald, 1964). This social network is the facilitator and broker of knowledge relating opportunities, transportation and employment in destination countries. McKenzie and Rapoport (2007) suggest a model of acceleration of migration followed by its decline. According to this model, migration may start due to any trigger, even random distribution. In most cases a migration emerge and then subsided, but in some cases (again, possibly a random process) it becomes a mass migration, which may continue for some time but ultimately will also subsided. The characteristics of the migration chain can be related to the dynamics of migration and if migration may then remain at a minor level or increase into a wave.

However there is a difference between relating to bilateral migration from the ethnographic perspective and the global network perspective (Davis et al., 2013; Fagiolo & Mastrorillo, 2013; Tranos et al., 2014). The global network approach is providing a complementary knowledge on the spatial pattern of migration and is offering a new terminology for migration phenomena for different migration strategies and the differences between source and destination countries.

Social systems can be seen as network of connections (Castels & Miller, 2008). A network is a concept of connections of various extents, from a single person's net of several agents, to a vast and growing network of millions. According to Graph Theory networks are measurable both quantitatively and qualitatively (Barabasi, 2002; Strogatz, 2001) as a system of nodes and links that may represent social, commercial, transportation, ecological or any other type of connection. Such flexibility enables the use of networks in life sciences and social sciences as well as in mathematics.

In the present work, we relate to a bilateral matrix of directed network of source and destination migration flows of most of the countries in the world. This approach is focus on the flow network and the difference between the source countries network and the destination countries network and the differences between regions. We use the network concepts of degree (the number of links) and strength/weight (the number of migrants per link) and analyze global and regional distribution of degree and weight in a directed migration network (Davis et al., 2013; Fagiolo & Mastrorillo, 2013; Tranos et al., 2014). Most publications on the subject deal exclusively with *im*migration, although the movement of people is, in fact, double, consisting of immigration and emigration. We consider and compare both these aspects of the phenomenon in all countries for which there are reliable data and statistics.

The novelty of our approach relatively to other network approaches stands in: 1). we took all the countries of the world trying to give the most general picture of the immigration phenomenon. 2) we consider the migration fluxes and not the stock of migrants which is a static quantity since immigration is dynamic.

## 3. Analysis of bilateral migration flows

### 3.1. Data base

A bilateral migration matrix (Persons et al., 2005; Ratha and Shaw, 2007) was published by the World Bank and updated in the Migration and Remittances Fact-book 2011, for the years 2006 & 2010. This data set can be accessed at these links[1]. In this database, migrants are defined as people who do not live where they were born (Ratha and Shaw, 2007). The matrix covers migration data on 226 countries and territories (source and destination) and is a unified data source for these countries, including: national census, population registration, national statistical bureaus and secondary sources for countries with limited migration information. The data were verified by world and regional migration models and techniques to assess missing data, and the database includes migrants according to current distribution, registered and un-registered, i.e. it is a mixture of heterogeneous data with disparities across countries, differences of definition, classification, and other lacunae, a collected sharing of data between countries. However, the resultant bilateral migration matrix is the fullest and up-to-date set of data, and is most useful for modeling world migration (Parsons et al., 2005; Ratha & Shaw, 2007).

---

[1] http://econ.worldbank.org/WBSITE/EXTERNAL/EXTDEC/EXTDECPROSPECTS/0,,contentMDK:22759429~pagePK:64165401~piPK:64165026~theSitePK:476883,00.html
http://econ.worldbank.org/WBSITE/EXTERNAL/EXTDEC/EXTDECPROSPECTS/0,,contentMDK:21154867~pagePK:64165401~piPK:64165026~theSitePK:476883,00.html

To estimate migration flows between the years 2006-2010, we subtracted the 2010 matrix from the 2006 matrix. This process yields a new matrix that indicates the flow of migrants during the four years period and the change in number of migrants in each country. Each cell in the new matrix represents the number of migrants who moved from country A to country B. The subtract action can provide positive and negative results. Positive number of migrants indicate a positive flow of migration from A to B, however negative number are much more difficult to indicate. Negative migrant number in a cell may indicate migrants whom return to their home country (and therefore are not migrants anymore) or migrants which moved to a third county C (in this case these migrants will be counted in the A to C cell). Therefore in the analysis of migration flow we considered only positive results and the approach to negative numbers will be developed in future studies. This resulted in a matrix of 206 source countries from which emigrants left, and 145 destination countries that received immigrants during this period (61 countries in this sample did not received immigrants). These two systems do not mirror one another since the network is directed and a country may receive migrants from a country but not send migrants back.

## 3.2 Global degree and weight statistics

We present statistics for degrees (D) and weight (W) for the destination and source countries, using the log-log Rank Size representation (Zipf) as well as histograms. Degree and weight are sorted by decreasing values, ranking R for a given size (S) or value. In Figure 1a we plotted logW vs. logR for destination, and in Figure 1b for source countries. There is some similarity between weight distributions for destination and source countries, and the histograms are also generalized with a large decrease for low values of W. The log-log plot fit of the source countries is expressed as: $y = 5.7(2.14 - x)^{0.15}$

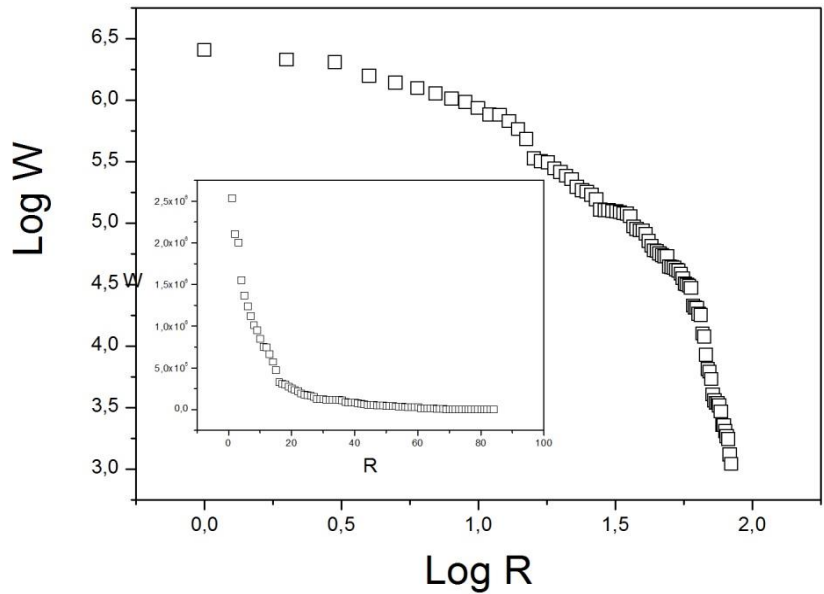

**Fig. 1a: W versus R (destination countries) in a log-log plot. The insert shows the curve W vs. R in linear plot to indicate the division into two parts**

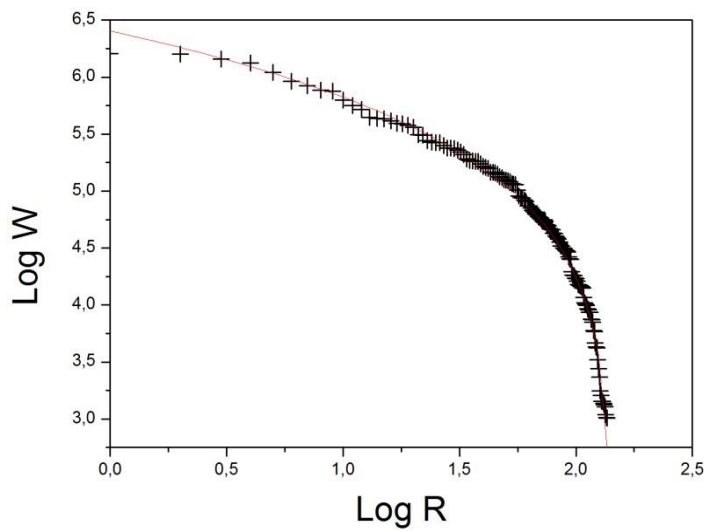

**Fig. 1b: W vs. R (for source countries) - log-log plot**

Closer inspection of the curves W(R) and LogW (LogR) of the destination countries shows a break at about W = 32,000 migrants. Although it is possible to fit the destination countries curve by summing two exponentials, this break can be interpreted as indicating separation between countries with W > 32000 and those with W < 32000.

From the same representation for the degrees (Figs.2a and 2b), there is a large difference in distributions of destination degrees and source degrees. The source distribution is homogeneous since all the countries are associated on the same curve, exception for some countries with higher values of D.

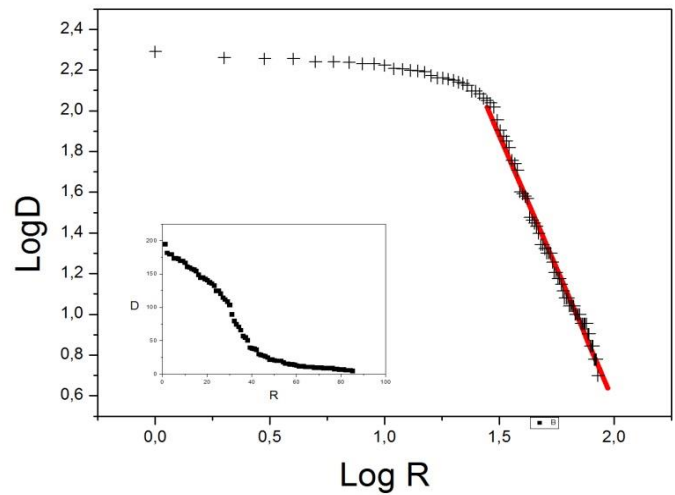

**Fig. 2a: D vs. R (destination countries) curve - log-log plot. The insert shows the curve D(R) in linear coordinates. There is a clear break in D(R) at about D = 100. For D < 100 the log-log curve is a straight line**

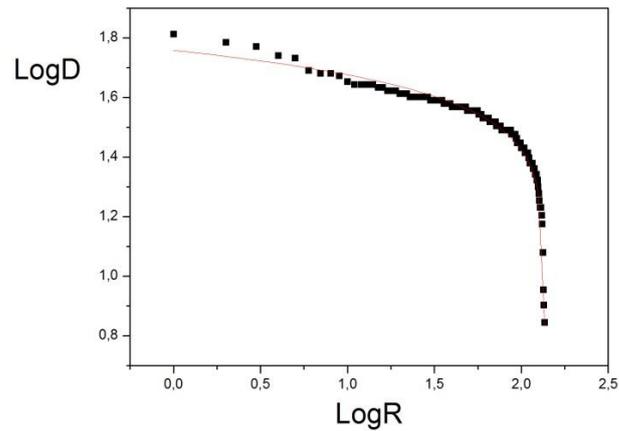

**Fig.2b: D vs. R (source countries) curve - log-log plot**

The histogram of degree distribution for the source countries shown in Fig.3 is very near a Gaussian distribution (see the continuous curve). The migration from source countries has a clear average value of 36 out links, and normal distribution around it, with a slight long right tail of highly connected source countries.

The log-log curve for destination countries is providing a different distribution and a compound of two parts, i.e. countries with degree larger than 100, and those with degree smaller than 100. For the latter, the Rank-Size curve is a straight line, i.e. LogD = 5.81 - 2.62 log R, or D is proportional to $R^{-2.62}$. This means that the distribution is a power law. The histogram of destination degrees (Fig.4) also shows division of the countries into two groups - those with D > 100 and those with D < 100.

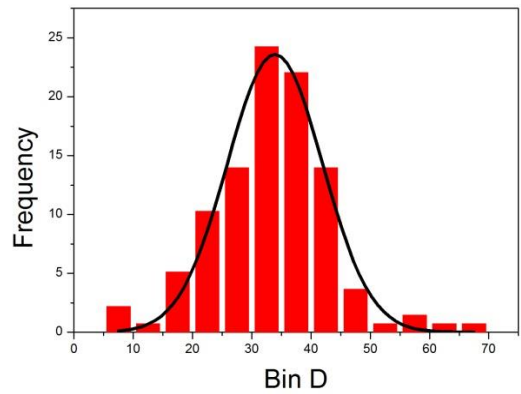

**Fig. 3: Histogram of degree distribution for the source countries. The continuous line is a Gaussian figure**

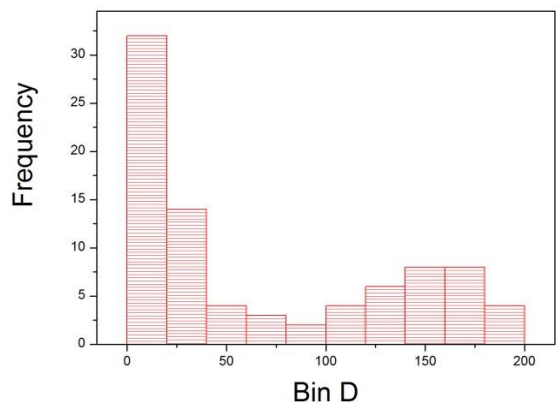

**Fig. 4: Histogram of degree distribution for destination countries, illustrating the two groups, those with low D and those with high D**

The difference between source and destination migration patterns requires further deepening, especially the differences in degree distribution. For source countries it seems that degree and weight are homogeneous, but for destination countries the situation is different. These findings have driven us to analyze the relationship of degree and weight and to finalize this analysis with regional characteristics.

## 3.3. Regional characteristics of destination countries

A log D vs Log R analysis of destination countries by regions, as seen in figure 5, is showing the different degree distributions in African countries, Latin America, Former Soviet union, Asia, Europe, New world and Middle East. There is a similar degree distribution of decline in most of the regions. However developed regions such as New World countries and European countries represent a different stage of network connectivity than African and Former Soviet Union countries. Latin American countries seem to be in an in between stage of a similar in migration strategy as developed countries, but the shape of the rest of the Latin American countries is decreasing faster than developed countries. Asian and Middle East countries are seem to be in an earlier migration stage of global connectivity. This analysis can be addresses as stages of in migration.

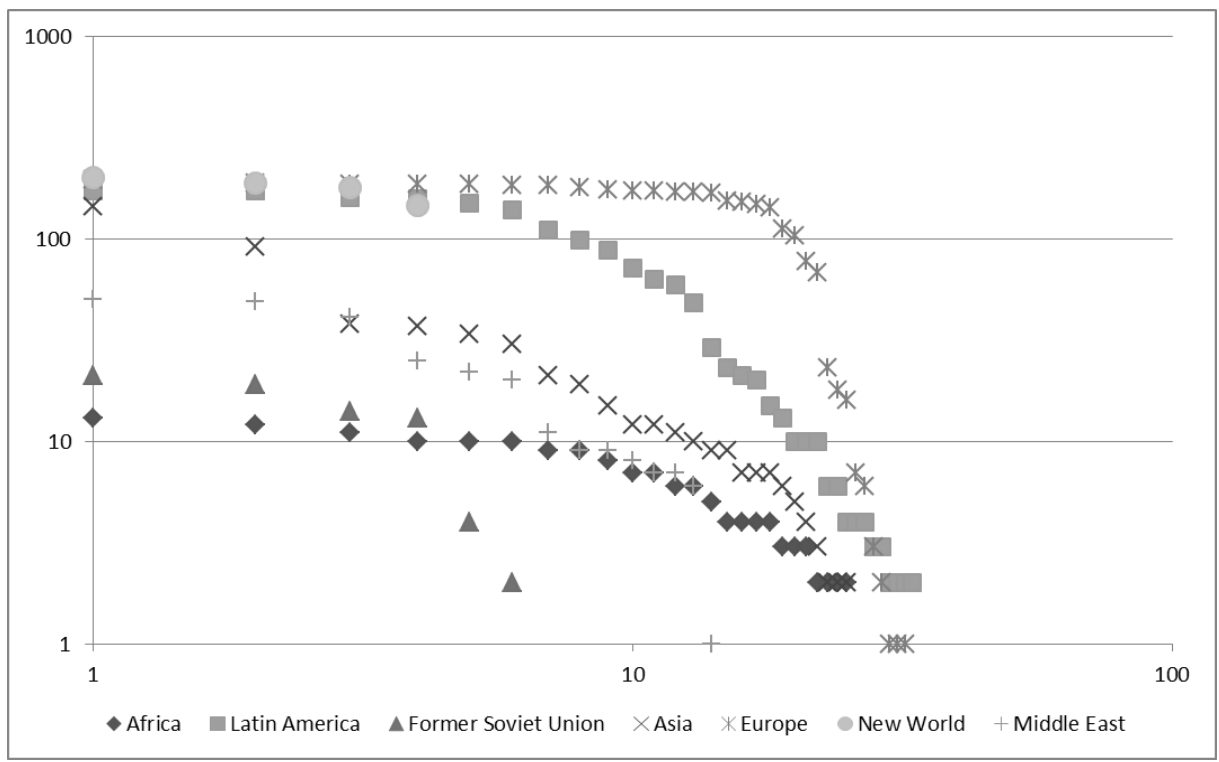

**Fig. 5: LogD vs LogR for destination countries, classified by regions**

These phenomena of migration strategy and migration stages have an addition angle in degree and weight analysis. The weight and degree parameters represent two aspects of destination countries' migration, namely the flow of migrants and the connectivity of the network. Figure 6 (D vs. W) shows clearly the groups of countries already observed in previous analysis i.e. those with large and those with small degree. We propose to classify countries in three groups. The group A (D > 100) includes most of the European countries, the New World (USA, Canada, Australia, New Zealand), and some Latin-American countries. The group B comprises countries with D < 100 and W > 32000. This group essentially includes African, Middle East and ex-Soviet Union countries. Finally the group C comprises countries with low D and low W, and includes mostly African, Asian and Latin-American countries.

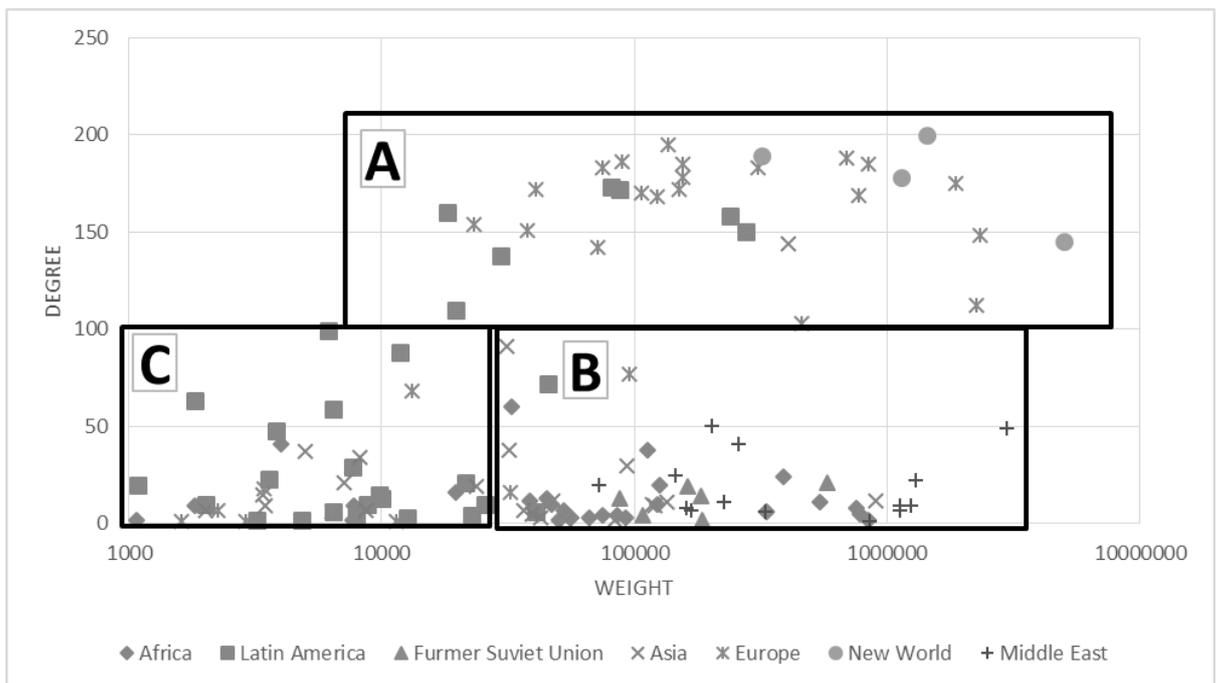

**Fig. 6: W vs. D for destination countries, classified in three groups according to values of W and D**

Destination countries are presenting a high level of differentiation also in the degree Vs average migration distance distribution (average of the in-links length between all source countries which send migrants to each destination country). Figure 7 represents the country degree and the average length of all its in-links in Km. There is a positive correlation between the degree and average in-links length. Good examples are the New World and European countries which have high degree and high average in-link length. But there are also more than few exceptions mainly of Asian and Latin Americans countries that have low degree and long average in-links length.

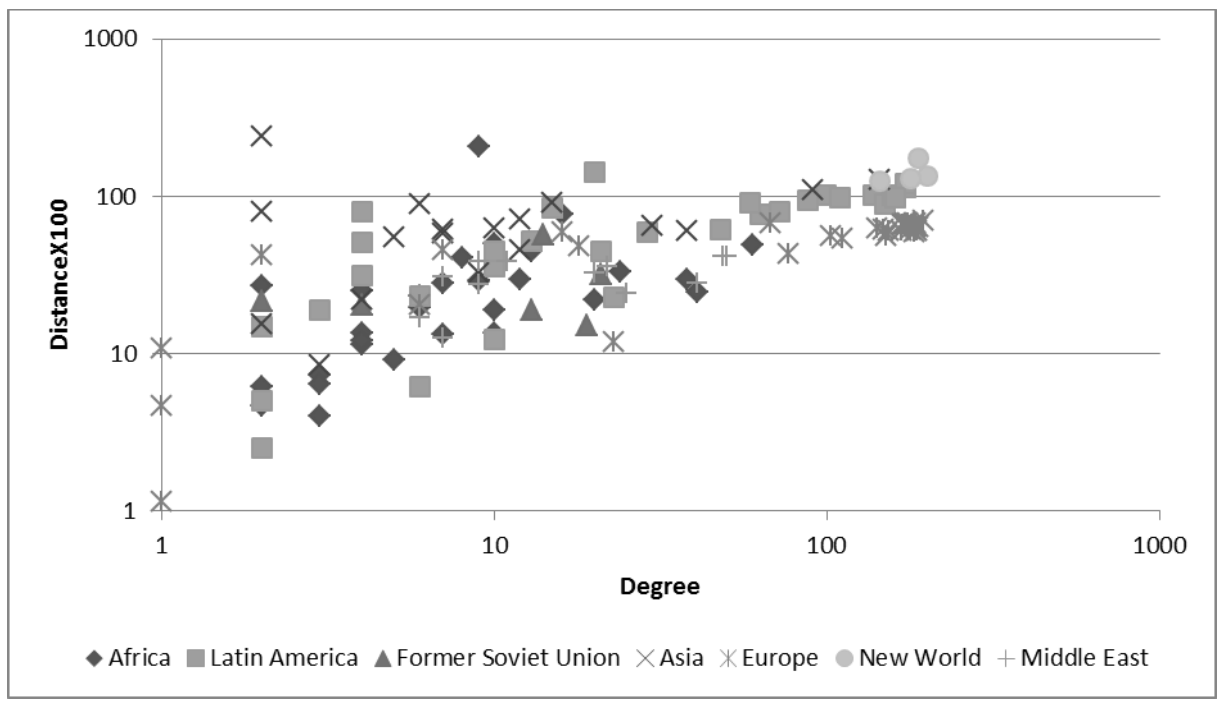

**Fig. 7: Degree vs. Distance for destination countries, classified by regions**

### 3.4 Regional characteristics of the source countries

Source countries are those from which migrants depart. According to migration data in 2006-2010, 206 countries experienced positive emigration, and are included in this analysis.

In opposition to the diversity of in-migration characteristics, out-migration from source countries is much more homogeneous as can be seen in figure 8. Degree ranking of regions displays a generalized decrease pattern for most of the countries, and it seems that there are only minimal differences of the out-migration pattern between regions and that all regions are in a similar migration stage.

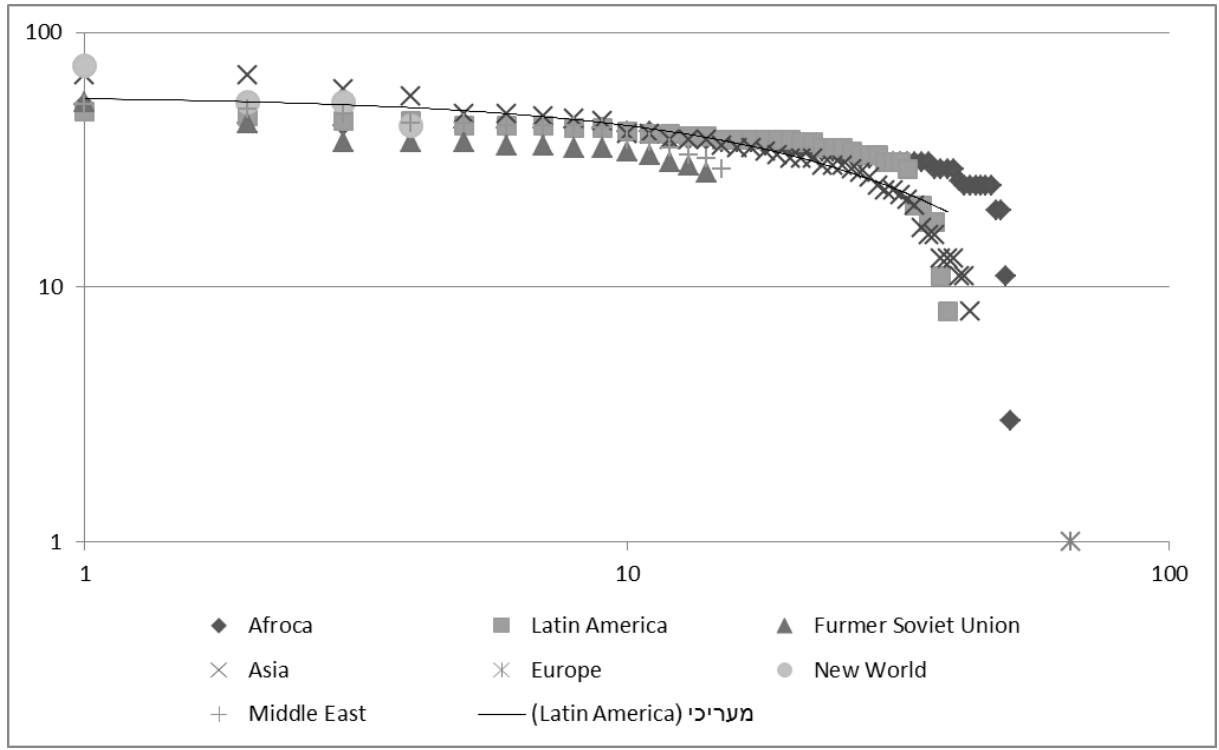

**Fig. 8: LogD vs LogR of source countries by regions**

Analysis of degree versus weight of source countries shows generally positive correlation and a totally different pattern from that of destination degree and weight (see figure 9). The figure also presents countries according to regions, but it is not possible to distinguish groups in this figure as we did for destination countries. This generalization of source countries and out migration has an important aspect in generalizing the migration phenomena on the one hand, and raises questions of the relationship of source and destination countries on the other hand.

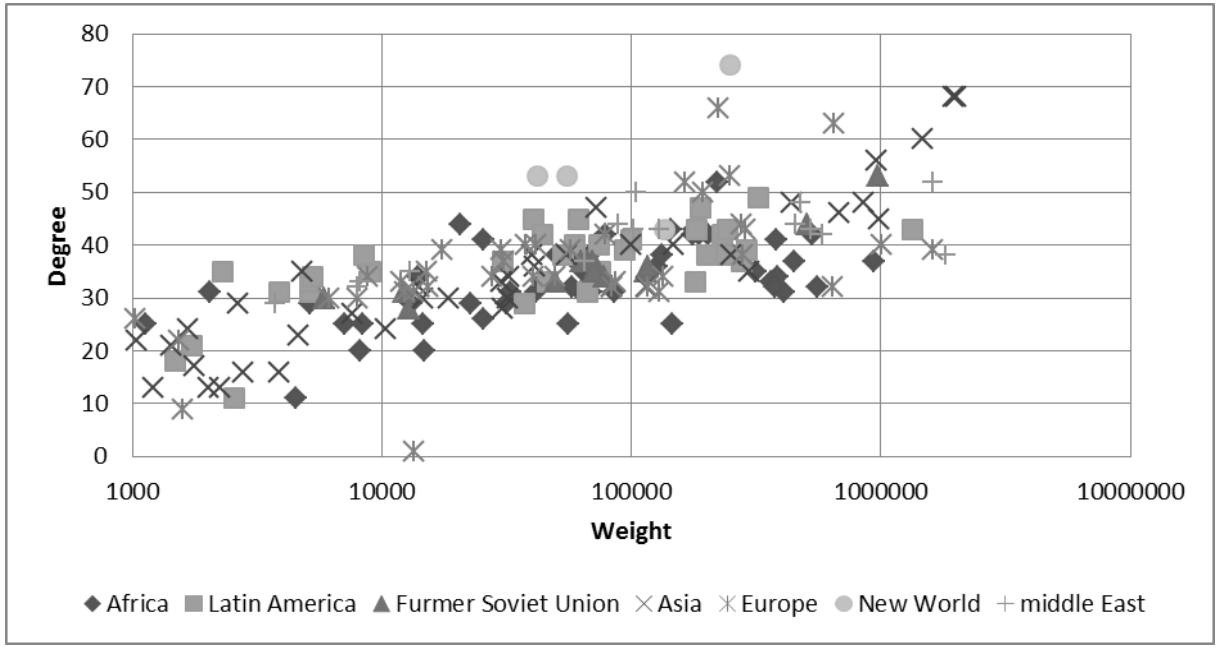

**Fig. 9: W versus D for source countries. Note the positive trend of the points for all the countries**

Source countries are presenting a low level of differentiation also in the degree Vs average migration distance distribution. Figure 10 represents the country degree and the average length of it's out-links in Km. There is an observed characteristic of average out-link length of 7800 Km for most of the countries and only few exceptions (S.D.=3100).

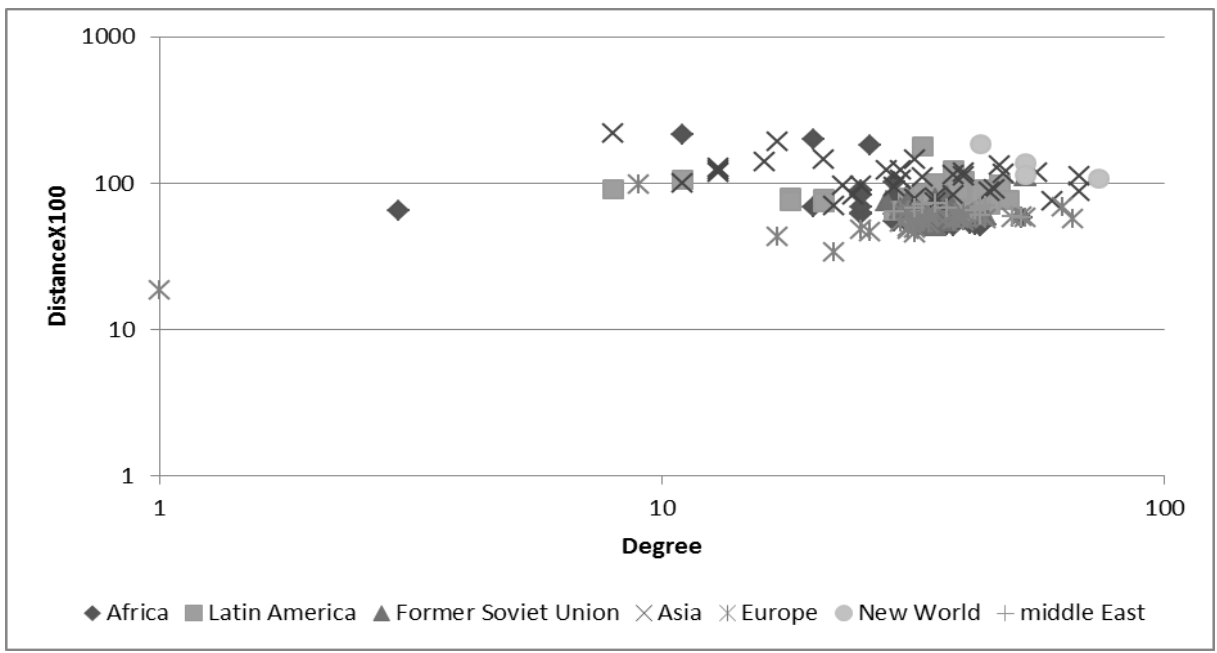

**Fig. 10: Degree vs. Distance for source countries, classified by regions**

# 4. Discussion

The concept of world migration degree and the analysis of directed migration flow in a bilateral network are providing an interesting perspective on the different migration systems of source and destination countries. From the perspective of source countries, out-migration is a well define phenomena with define characteristics of out-links. It is remarkable that the degree distribution of source countries is Gaussian, indicating some universality in the emigration phenomenon. The motivations for people to leave their country vary from country to country and from one social class to another etc, but on the other hand the pattern of leaving is quite homogenous. Source countries have an average of 34 out-links (s.d.=11) a generalized distribution of out-links in a ranking graph, and an average out-migration distance of 7800 Km. (s.d.= 3100Km), source countries also have a positive correlation between the degree and the weight. These characteristics are relating to most of the countries in this analysis (N=206) and these countries cannot be separated to regions. Migration degree looks at the out-migration as general phenomenon, and most of the countries have similar characteristics.

The perspective of in-migration is quite different, and sub-groups can be defined. Firstly there are two define groups of high and low network connectivity. Countries which have 100 in-degree and above have a different (much less declining) pattern in a ranking graph than countries which have less than 100 in-degree. Secondly the in-migration can be defined by regions, were new world and European countries are highly connected to the global migration network with high degree, high weight and long distanced links; and African and Asian countries have much less connectivity to global migration network with low degree, low weight and much sorter links. Between these two edges of migration connectivity we can also define two mid-groups of destination migration countries: Latin American countries have low weight, but some of them have high degree and long distance links; on the other hand, former Soviet Union and Middle Eastern countries have low degree and short links, but high weight. These mid groups represent two different strategies for increasing migration: to increase connectivity and increase the degree vs increase the amount of migrants in existing links and increase total migration weight.

We can also look at these strategies from the south-south perspective and separate the migration ports of high weight and low degree as Qatar, United Arab Emirates and Tajikistan; from migration hubs of high connectivity such as Latin American countries and also Central African Republic, Sudan and Mauritania.

## 5. Conclusions

We have proposed a new approach to typologies of the world migration phenomenon as a complex network, and have tried to compose a general picture by evaluating both countries as sources of migrants and countries as receivers of migrants. We employed two parameters: degree (D) - the number of links of a country with the rest of the world; and weight (W) - the number of migrants to and from a country and analyzed the average links distance. The weights and degrees are completely different for each migration system. In particular, source countries present a generalized phenomenon with common pattern of positive correlation of population with degree and weight. Destination countries fall into three groups according to degree, weight and migration distance: developed countries (European and new world countries) are highly connected to the migration network and represent a high degree, high weight and long distanced links; African and Asian countries represents a low degree, low weight and; former Soviet union and Middle Easters countries represent a low degree, short distance links and high weight; and Latin American countries represent a high degree, long distance links and low weight.

Migration literature focuses mainly on characteristics of strong links, because these links are largely responsible for migration flows. However, degree also has value in the analysis of migration as a global network, and may characterize the level of connectivity of a country to the migration network and define migration typologies. It is also important for analysis of migration dynamics (the subject of a future research) either as different phenomena of relationships between countries, or as the same phenomenon at different stages.

# References

Alterman, R. (2002) *Planning in the face of crises*, Routledge

Barabási, A.L. (2002) *Links: The New Science of Networks* (Perseus, Cambridge, MA)

Castles, S. (2010) Understanding Global Migration: A Social Transformation Perspective, *Journal of Ethnic and Migration Studies*, **36**, (10) 1565-1586

Castles, S. Miller, M. J. (2008) *The Age of Migration*, 4th revised and updated edition (1998), Palgrave MacMillan (UK) and Guildford Press (US)

Cohen, R. (1995) *The Cambridge Survey of World Migration*, University of Cambridge

De Haas, H. (2010a) Migration and Development: A theoretical perspective, *International Migration Review* 44 (1) 227-264

De Hass, H. (2010b) The Internal Dynamics of Migration Processes: A Theoretical Inquiry, *Journal of Ethnic and Migration Studies* **36** (10) 1587-1617

Durand, J. Massey, D. S. (2010) New World Orders: Continuities and Changes in Latin American Migration, *The Annals of the American Academy of Political and Social Science* **630** (1) 20-52

Fagiolo, G. (2013) International migration network: Topology and modeling, *Physical Review E* 88, 012812

Goldin, I. Cameron, G. Balarajan, M. (2011) *Exceptional People* (Princeton University Press,

Hatton, T.J. Williamson, J.G. (2005) *Global Migration and World economy: two centuries of Policy and Performance*, Cambridge University Press

Koser, K. Laczko, F. editors (2010) World Migration Report 2010 – The Future of Migration: Building Capacities for Change, Geneva: International Organization for Migration.

Kritz, M.M. Lim, L.L. Zlotnik, H. (1992) International migration systems: a global approach, **12** 354

Li, P.S. (2008) World Migration in the Age of Globalization: Policy Implications and Challenges, *New Zealand Population Review* **33/34** 1-22

MacDonald, J.S. MacDonald, L.D. (1964) Chain migration ethnic neighborhood Formation and social networks, *The Milbank Memorial Fund Quarterly*, **42**(1) 82 -97

Massey, D.S. (1990) The Social and Economic Origins of Immigration, *Annals of the American Academy of Political and Social Science* **510** 60 -72

Massey, D.S. Arango, J. Hugo, G. Kouaouci, A. Pellegrino, A. Taylor, J.E. (1993) Theories of International Migration: A Review and Appraisal, *Population and Development Review* **19** (3) 431-466

Massey, D.S. Goldring, L. Durand, J. (1994) Continuities in Transnational Migration: An Analysis of Nineteen Mexican Communities, *The American Journal of Sociology* **99**1492-1533

Massey, D.S. Taylor, J.E. (2004) *International Migration: Prospects and Policies in a Global Market*, Oxford University Press

McKenzie, D. Rapoport, H. (2007) Network Effects and the Dynamics of Migration and Inequality: Theory and Evidence from Mexico", *Journal of Development Economics* **84**1-24

Nasra, M. Shah, I. M, (1999) Chain Migration Through the Social Network: Experience of Labor Migrants in Kuwait, *International Migration*, **37**(2) 361–382,

Nijkamp, P. Gheasi, M. Rietveld, P. (2011) Migrants and international economic linkages: a metaoverview, *Spatial Economic Analysis* *6(4)* 359-376

Nijkamp, P. Poot, J. Sahin, M. (2012) *Migration Impact Assessment: New Horizons,* Edward Elgar, Cheltenham, Glos

Okruhlik, G. Conge, P. (1997) National Autonomy, Labor Migration and Political Crisis: Yemen and Saudi Arabia, *Middle East Journal*, **51** (4) 554–565

Persons, R.C. Skeldon, R. Walmsley, T.L. Winters, L.A. (2005) Quantifying the international bilateral movements of migrants, Globalization and poverty working paper T13, Development Research Center on Migration University of Sussex

Rainer, M. (2007) Migration, labor markets, and integration of migrants: An overview for Europe, HWWI Policy Paper, No. 3-6

Ratha, D. Shaw, W. (2007) South-South migration and remittances, World Bank Working Paper No. 102

Sandell, R. (2006) Spain's immigration experience: lessons to be learned from looking at the statistics, Working Paper 30, Real Instituto Elcano

Shah, N.M. Menon, I. (1999) Chain migration through the social network: experience of labor migrants in Kuwait, *International Migration* **37** (2) 361-382

Strogatz, S. (2001) Exploring complex networks, *Nature* **410**(6825) 268-276

Taylor, J.E. (1999) The new economics of labor migration and the role of remittances in the migration process. *International Migration* **37** 63–86.

Tranos, E. Gheasi, M. Nijkamp, P. (2014) International migration: a global complex network, *Environment and Planning B: Planning and Design* 2014, **41** online

Zipf, G.K. (1949) *Human Behavior and the Principle of Least Effort,* Addison-Wesley, New York

Zolbert, A.R. (1989) The Next Waves: Migration Theory for a Changing World, *International Migration Review*, **23** 403-430